\documentclass[aps,pra,twocolumn,superscriptaddress]{revtex4}

\usepackage{color}
\usepackage{amssymb}
\usepackage{amsmath, amsthm}
\usepackage{epsfig}
\usepackage{epstopdf}
\usepackage{graphicx}
\usepackage{notoccite}

\newcommand{\I}{\mathbb I}

\newcommand{\ket}[1]{\vert#1\rangle}
\newcommand{\bra}[1]{\langle#1\vert}

\usepackage{bm}

\newcommand{\ba}{\begin{eqnarray}}
\newcommand{\be}{\begin{equation}}
\newcommand{\ee}{\end{equation}}
\newcommand{\beq}{\begin{equation}}
\newcommand{\eeq}{  \end{equation}}
\newcommand{\bea}{\begin{eqnarray}}
\newcommand{\eea}{  \end{eqnarray}}
\newcommand{\ea}{\end{eqnarray}}
\newcommand{\ban}{\begin{eqnarray*}}
\newcommand{\ean}{\end{eqnarray*}}

\newcommand{\tr}{\operatorname{Tr}}

\newcommand{\ie}{{\it{i.e.}~}}

\begin{document}

\title{Experimental nonlocality-based randomness generation with non-projective measurements}

\begin{abstract}
We report on an optical setup generating more than one bit of randomness from one entangled bit (\ie a maximally entangled state of two-qubits). The amount of randomness is certified through the observation of Bell non-local correlations. To attain this result we implemented a high-purity entanglement source and a non-projective three-outcome measurement. Our implementation achieves a gain of 27\% of randomness as compared with the standard methods using projective measurements. Additionally we estimate the amount of randomness certified in a one-sided device independent scenario, through the observation of EPR steering. Our results prove that non-projective quantum measurements allows extending the limits for nonlocality-based certified randomness generation using current technology.
\end{abstract}


\author{S. G\'omez}
\affiliation{Departamento de F\'isica, Universidad de Concepci\'on, 160-C Concepci\'on, Chile}
\affiliation{Center  for  Optics  and  Photonics,  Universidad  de  Concepci\'on,  160-C  Concepci\'on,
Chile}
\affiliation{MSI-Nucleus  for  Advanced  Optics,  Universidad  de  Concepci\'on,  160-C  Concepci\'on,
Chile}

\author{A. Mattar}
\affiliation{ICFO-Institut de Ciencies Fotoniques, The Barcelona Institute of Science and Technology, 08860 Castelldefels, Barcelona, Spain}

\author{E. S. G\'omez}
\affiliation{Departamento de F\'isica, Universidad de Concepci\'on, 160-C Concepci\'on, Chile}
\affiliation{Center  for  Optics  and  Photonics,  Universidad  de  Concepci\'on,  160-C  Concepci\'on,
Chile}
\affiliation{MSI-Nucleus  for  Advanced  Optics,  Universidad  de  Concepci\'on,  160-C  Concepci\'on,
Chile}

\author{D. Cavalcanti}
\affiliation{ICFO-Institut de Ciencies Fotoniques, The Barcelona Institute of Science and Technology, 08860 Castelldefels, Barcelona, Spain}

\author{O. Jim\'enez Far\'ias}
\affiliation{ICFO-Institut de Ciencies Fotoniques, The Barcelona Institute of Science and Technology, 08860 Castelldefels, Barcelona, Spain}
\author{A. Ac\'in}
\affiliation{ICFO-Institut de Ciencies Fotoniques, The Barcelona Institute of Science and Technology, 08860 Castelldefels, Barcelona, Spain}
\affiliation{ICREA-Instituci\'o Catalana de Recerca i Estudis Avan\c cats, Lluis Companys 23
Barcelona, 08010, Spain}
\author{G. Lima}
\affiliation{Departamento de F\'isica, Universidad de Concepci\'on, 160-C Concepci\'on, Chile}
\affiliation{Center  for  Optics  and  Photonics,  Universidad  de  Concepci\'on,  160-C  Concepci\'on,
Chile}
\affiliation{MSI-Nucleus  for  Advanced  Optics,  Universidad  de  Concepci\'on,  160-C  Concepci\'on,
Chile}


\date{\today}

\pacs{}

\maketitle

The existence of random processes, besides having philosophical consequences, has applications in many disciplines such as cryptography and simulations of physical, biological, and social phenomena. Mismatches between the modelling and the actual working of random number generators (RNGs) may lead to wrong conclusions. Quantum technologies provide a solution to this problem through device-independent (DI) randomness generation protocols \cite{colbeck06r,Pironio:2010kx,acin16} built from Bell nonlocal correlations \cite{Bell_original,Brunner:2014ix}. To date, all implementations of DIRNGs used projective measurements on quantum bits \cite{Pironio:2010kx,Kwiat13,nist17}, thus being limited to one random bit per round and particle. Here, we report on an optical setup providing more than one random bit per round from one entangled bit \cite{Acin:2016cl}. To attain this result we implement a Bell test involving a non-projective measurement on an entangled state of high purity. Our work demonstrates the importance of non-projective measurements to attain the ultimate limits for DIRNG.

\begin{figure}
\centering
\includegraphics[width = .95\linewidth]{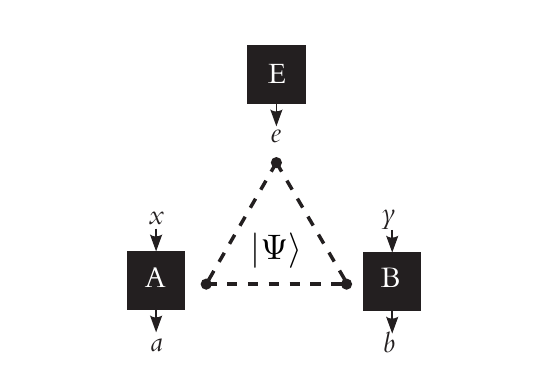}
\caption[]{Device-independent randomness generation scenario: a user applies uncharacterised measurements $x$ and $y$ to two devices A and B, obtaining outcomes $a$ and $b$ respectively. In the experiment, the user assumes that the state of the two devices A and B is the reduced state of a pure tripartite quantum state $\ket{\Psi}$ correlated with an adversarial party, Eve, who holds device $E$. No further assumption is made on Eve, who could have a complete description not only of this quantum state, but also of all the measurements performed on it. In order to guess the outcomes produced in the experiment, Eve applies a measurement to her device $E$ that produces outcomes $e$. Without loss of generality, this outcome can be seen as Eve's guess on the user's results.}
\label{fig:scenario}
\end{figure}

The standard scenario for non-locality-based randomness generation consists of a user, who has access to two quantum measurement devices, A and B, which have input choices and provide outputs \cite{acin16}, see Fig.\ref{fig:scenario}. The user's goal is to certify that the outcomes produced in the experiment are random. We consider the strongest definition of randomness in which the user's outcomes are demanded to be unpredictable not only to her, but to any other observer \cite{acin16}. This, besides being fundamentally important, guarantees that the obtained randomness is private, a requirement for cryptographic applications \cite{colbeck06r,Pironio:2010kx,CK11}. In the device-independent scenario nothing is assumed on the inner working of the measurement devices, which are treated as quantum black boxes fed with classical inputs $x$ and $y$ - the measurement choices - and producing classical outputs $a$ and $b$ - the measurement results. After collecting enough statistics, the user's description of the devices is given by the set of conditional probabilities $P(ab|xy)$.

In randomness certification protocols it is assumed that the AB state is the reduced state of a tripartite state $\ket{\Psi}_{ABE}$ produced by an outsider, Eve, who holds a device E. Moreover, Eve could have prepared the measurement devices, and thus has a complete description of the measurements in A and B. The randomness in user's outcome $a$ for a particular measurement $x=x^*$ can be estimated through the so-called guessing probability \cite{NPS14,BSS14}:
\begin{align}
\label{eq:randomnessSDP}
&P_{\text{guess}}=\max_{\{\ket{\Psi},\Pi_{a|x},\Pi_{b|y},\Pi_{e}\}} \sum_a\bra{\Psi} \Pi_{a|x^*}\otimes\I\otimes\Pi_{e=a}\ket{\Psi}\\
&\textrm{such that}\nonumber\\
& P(ab|xy)=\bra{\Psi} \Pi_{a|x}\otimes\Pi_{b|y}\otimes\I\ket{\Psi}.
\end{align}
This quantity gives the maximum probability that E's outcome $e$ matches the user's outcome $a$ for measurement $x^*$ over all possible quantum realisations, described by a tripartite quantum state $\ket{\Psi}$ and measurements $\Pi_{a|x}$, $\Pi_{b|y}$ and $\Pi_e$ for devices A, B and E, compatible with the observed distribution $P(ab|xy)$. The guessing probability can be upper bounded by semi-definite programming (SDP) techniques \cite{NPS14,BSS14}. The estimated randomness can be expressed in bits through $R=-\log_2 (P_{\text{guess}})$.
In order to guarantee some amount of randomness the user's observed correlations must be nonlocal, that is, violate a Bell inequality. If this is not the case, they can be reproduced by a local and deterministic model and therefore $P_\text{guess}=1$ \cite{acin16}.

The main motivation of this work is to probe the ultimate limits for randomness certification using quantum resources. In order to observe a Bell violation between A and B, the user's state must be entangled. If the state is of two qubits and the measurements are projective, as in standard Bell experiments, one cannot certify more than one random bit from each qubit. However, this is no longer the case if one uses non-projective measurements \cite{Acin:2016cl}.  We report here a photonic experiment demonstrating how non-projective measurements offer a significant advantage in a Bell scenario and allow one to certify more than one random bit from a qubit.

\begin{figure*}
\centering
\includegraphics[width = 1\linewidth]{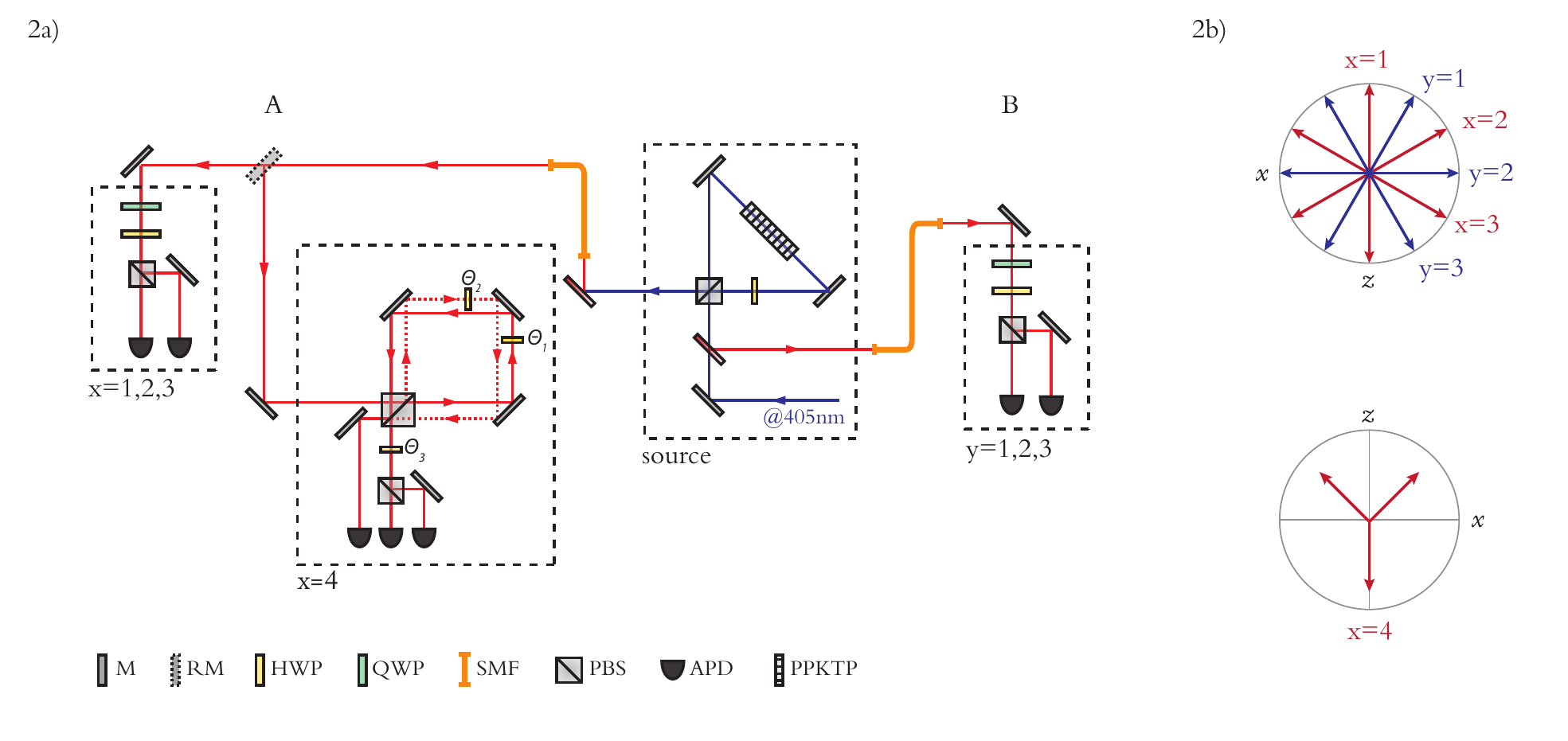}
 \caption{(Color Online) a) Our experimental setup is composed by an ultra-bright parametric down-conversion source (see main text) generating a near-perfect $|\psi^-\rangle=(|HV\rangle-|VH\rangle)/\sqrt{2}$ polarization state, followed by polarization measurements in each photon. The measurements $x,y=1,2,3$ have binary outcomes and are implemented using a quarter-wave plate (QWP), a HWP and a PBS, followed by avalanche photodiode detectors (APD). A removable mirror (RM) allows to select between measurements $x=1,2,3$ and $x=4$. The fourth and non-projective measurement, $x=4$, performed by device A is implemented by a double path Sagnac-interferometer. b) Bloch sphere representation of the measurements performed in A and B.  The measurements labeled by $x,y=1,2,3$ are given by symmetrically spaced two-outcome (projective) measurements in the x-z plane, and correspond to the settings required to maximally violate the chained Bell inequality \cite{chained}. Measurement $x=4$ has three outcomes, corresponding to Bloch vectors equally spaced in the x-z plane.}

\label{fig:setup}
\end{figure*}

Our experiment is similar to a standard Bell test using photons entangled in polarization (see Figure~\ref{fig:setup}a). However we need to solve two experimental challenges that make it unique with respect to previous experiments and that are crucial to achieve the certification of more than one random bit. First, we need to prepare a highly entangled state providing a very high two-photon visibility. To achieve this, we use an ultra-bright spontaneous parametric down-conversion source, where a type-II non-linear periodically poled potassium titanyl phosphate (PPKTP) crystal is pumped by a continuous wave 405~nm laser to generate 810~nm polarization-entangled photons \cite{Kim2006a,Kim2006b,Zeilinger2007,Ljunggren_PRA_fibers}. The non-linear crystal is placed inside an intrinsically phase-stable Sagnac interferometer, which is composed of two laser mirrors, a half-wave plate (HWP), and a polarizing beamsplitter (PBS) cube. The clockwise and counter-clockwise propagating modes of the generated pair of photons overlap inside the interferometer resulting in the bi-photon Bell state $|\psi^-\rangle=(|HV\rangle-|VH\rangle)/\sqrt{2}$. We carefully control the spatial and spectral modes of the generated photons. Semrock high-quality (peak transmission $>$ $90\%$) narrow bandpass (FWHM -full width at half maximum - of 0.5 nm) filters centered at $810$ nm are used to ensure that phase-matching conditions are achieved with the horizontal and vertical polarization modes at degenerated frequencies. Then, we enforce path indistinguishability of the photon pair modes (``HV'' and ``VH'') by coupling the generated down-converted photons into single mode fibers (SMF) after being transmitted by the PBS. We also adopt high-quality polarizing optics components to ensure a polarization extinct ratio greater than $10^7$:1. This guarantees that the two-photon visibility is not limited by the polarization contrast of the detection apparatuses. Last, we use a high-resolution coincidence field programmable gate array electronics to implement 500~ps coincidence windows, thus drastically reducing the accidental coincidence count probability to less than $10^{-5}$ (PerkinElmer single-photon
avalanche detectors with an overall detection efficiency of $15\%$ were used). Thanks to these measures, we attain a high overall two-photon visibility of ($99.7 \pm 0.2$)\%.

Second, and contrary to standard Bell tests, our experiment consists not only of projective measurements, but involves a non-projective measurement defined by a Positive-Operator-Valued Measure (POVM). Indeed, while device B applies 3 projective measurements (labeled by $y$=1,2,3), device A can implement 3 projective measurements ($x$=1,2,3) plus a POVM measurement ($x$=4) of three outcomes. All projective measurements are implemented by usual polarization analyzers.
The non-projective measurement used in our experiment consists of three outcomes, associated with POVM elements $\Pi_i=\frac{2}{3}\ket{\psi_i}\bra{\psi_i}$, where
\begin{eqnarray}
&&\ket{\psi_0}=\ket{V}, \nonumber\\
&&\ket{\psi_{1}}=-\frac{1}{2} (\ket{V}+\sqrt{3}\ket{H}),\\ \nonumber
&&\ket{\psi_{2}}=-\frac{1}{2} (\ket{V}-\sqrt{3}\ket{H}).
\end{eqnarray}
This measurement is obtained in our setup by coherently coupling the polarisation of photons with additional spatial modes in a a double-path Sagnac interferometer (see Fig. 2a), similar to the one reported in \cite{gomez2016device}. The propagation modes of a photon within this interferometer are not co-propagating and depend on the polarization state, which allows for conditional polarization transformation through HWPs at angles $\theta_1=0$ and $\theta_2=2\sin^{-1}(\sqrt{2/3})$. These two propagation modes are then superposed again in the PBS generating two new outcome modes. The first is sent directly to the detector (first outcome), while the other experience an additional HWP with $\theta_3=\pi/2$ and is split by a PBS to generate two modes representing the second and third outcomes of the non-projective measurement. The Bloch sphere representation of these measurements are shown in Fig. 2b.

We also notice that in our work we invoke the fair-sampling assumption \cite{Brunner:2014ix}, which we use to discard the no-detection events. This assumption is highly debatable in DI cryptographic applications, in which two distant users are connected by a channel whose losses can be simulated by an eavesdropper. But note that it is less critical in DIRNGs protocols in which the two devices are in the same location and under control of a honest user.

Using the estimated visibilities we first run a numerical search to find measurements that maximise the amount of randomness generated in our scenario. This search led us to the measurement settings shown in Fig. 2b. By implementing these measurements we obtained a collection of observed experimental frequencies $f(ab|xy)$ that we use to estimate randomness in the following way (see Supp. Mat. for more details). The raw data obtained from measurements is avilable in \cite{gitrep}. We first used a regularization method \cite{regularization} based on the Collins-Gisin parametrization \cite{CollinsGisin} of the space of probabilities to generate a set of no-signaling probability distributions $P_{NS}=\{P_{NS}(ab|xy)\}$. By considering marginal probabilities $P(a|x,y=1)$ and $P(b|x=1,y)$, the Collins-Gisin representation enforces no-signaling constraints of $P$ by dropping all probabilities involving the last outcome of all measurements. For instance, for the POVM that has 3 outcomes, the probability $P_{NS}(a=3,b|xy)$ is implicitly set by imposing $P_{NS}(a=3,b|xy) = P(a=3|x,y=1) - P(a=2,b|xy) - P(a=1|xy)$.

With $P_{NS}$ we run the semidefinite (SDP) program proposed in Refs. \cite{NPS14,BSS14}, that provides an upper bound to the guessing probability \eqref{eq:randomnessSDP}.
The solution of this SDP optimisation provides a linear function $S(P)$ whose value is a lower bound on the amount of randomness of any set of distributions $P$. We finally rewrite $S$ in terms of expected values and use it to estimate the amount of randomness in our experiment. The errors of the recorded probabilities are calculated assuming fair samples from Poissonian distributions and Gaussian error propagation. Notice that our statistical analysis is done under the assumption of independent and identical distributed (iid) copies of the same experiment, for which the guessing probability ~\eqref{eq:randomnessSDP} is well defined. Other statistical methods are available \cite{Pironio:2010kx,zhang11y,nist17}, but this is beyond the scope of this work.

After these steps we were able to certify
\begin{equation}
R_{povm}^{DI}=1.18\pm{0.08}
 \end{equation}
bits of randomness per use of the devices. As a matter of comparison we also performed the same analysis in the case Alice and Bob use only the projective measurements $x,y=1,2,3$ and randomness is obtained from the setting $x=1$. In this case $R_{proj}^{DI}=0.93 \pm 0.08$. Thus, the addition of a 3-outcome non-projective measurement provided a gain of 27\% of randomness.

In our setup, we can also certify randomness in a semi-device independent scenario in which device B is assumed to be fully characterised. In this scenario randomness can be certified by the presence of quantum steering \cite{WJD07}, a situation where the box A is still treated as black boxes with inputs $x$ and outputs $a$, while B is assumed to be able to make tomography of the conditional states $\rho_{a|x}^B$. The information the user has in this situation can be summarized in the set of unnormalized quantum sates  $\{\sigma_{a|x}^B\}_{a,x}$, where $\sigma_{a|x}^B=p(a|x)\rho_{a|x}^B$. Notice that, in order to obtain a set $\{\sigma_{a|x}^B\}_{a,x}$ that satisfy the no-signalling conditions $\sum_a \sigma_{a|x}^B=\sum_a \sigma_{a|x'}^B$ we also need to resort on the distributions $P_{NS}$ obtained through the Collins-Gisin parametrization. Given the knowledge of $\{\sigma_{a|x}^B\}_{a,x}$, the guessing probability of the outcome $a$ of the measurement $x'$ can also be estimated through a semi-definite program \cite{PCSA15}. The amount of randomness can be calculated in a similar manner as in \eqref{eq:randomnessSDP} \cite{PCSA15} and, in this case, we were able to certify $R^{St}_{povm}=1.27\pm0.14$.

In the context of certified RNG protocols, our work is relevant both from a fundamental and applied perspective, as it demonstrates how the more general class of non-projective quantum measurements allows extending the limits for nonlocality-based certified randomness generation using current technology. In the scenario of device-independent quantum information processing, we show that a gain of 27$\%$ in the rate of random bit string generation is possible. In the case of semi-device independent RNG protocols, we demonstrate that this gain can be improved to $36\%$.

We thank A. Cabello, T. V\'ertesi, and M. Pawlowski for discussions. This work was supported by the Ram\'on y Cajal fellowship, Spanish MINECO (QIBEQI FIS2016-80773-P and Severo Ochoa SEV-2015-0522), the AXA Chair in Quantum Information Science, Generalitat de Catalunya (SGR875 and CERCA Programme), Fundaci\'{o} Privada Cellex, Fondecyt~1160400, CONICYT~PFB08-024,and Milenio~RC130001. E.S.G. acknowledges support from Fondecyt~11150325. S.G. acknowledges CONICYT.


\begin{thebibliography}{23}
\expandafter\ifx\csname natexlab\endcsname\relax\def\natexlab#1{#1}\fi
\expandafter\ifx\csname bibnamefont\endcsname\relax
  \def\bibnamefont#1{#1}\fi
\expandafter\ifx\csname bibfnamefont\endcsname\relax
  \def\bibfnamefont#1{#1}\fi
\expandafter\ifx\csname citenamefont\endcsname\relax
  \def\citenamefont#1{#1}\fi
\expandafter\ifx\csname url\endcsname\relax
  \def\url#1{\texttt{#1}}\fi
\expandafter\ifx\csname urlprefix\endcsname\relax\def\urlprefix{URL }\fi
\providecommand{\bibinfo}[2]{#2}
\providecommand{\eprint}[2][]{\url{#2}}

\bibitem[{\citenamefont{Colbeck}(2006)}]{colbeck06r}
\bibinfo{author}{\bibfnamefont{R.}~\bibnamefont{Colbeck}}, Ph.D. thesis,
  \bibinfo{school}{University of Cambridge} (\bibinfo{year}{2006}).

\bibitem[{\citenamefont{Pironio et~al.}(2010)\citenamefont{Pironio, Ac{\'i}n,
  Massar, de~la Giroday, Matsukevich, Maunz, Olmschenk, Hayes, Luo, Manning
  et~al.}}]{Pironio:2010kx}
\bibinfo{author}{\bibfnamefont{S.}~\bibnamefont{Pironio}},
  \bibinfo{author}{\bibfnamefont{A.}~\bibnamefont{Ac{\'i}n}},
  \bibinfo{author}{\bibfnamefont{S.}~\bibnamefont{Massar}},
  \bibinfo{author}{\bibfnamefont{B.}~\bibnamefont{de~la Giroday}},
  \bibinfo{author}{\bibfnamefont{D.}~\bibnamefont{Matsukevich}},
  \bibinfo{author}{\bibfnamefont{P.}~\bibnamefont{Maunz}},
  \bibinfo{author}{\bibfnamefont{S.}~\bibnamefont{Olmschenk}},
  \bibinfo{author}{\bibfnamefont{D.}~\bibnamefont{Hayes}},
  \bibinfo{author}{\bibfnamefont{L.}~\bibnamefont{Luo}},
  \bibinfo{author}{\bibfnamefont{T.}~\bibnamefont{Manning}},
  \bibnamefont{et~al.}, \bibinfo{journal}{Nature}
  \textbf{\bibinfo{volume}{464}}, \bibinfo{pages}{1021} (\bibinfo{year}{2010}),
  ISSN \bibinfo{issn}{0028-0836}.

\bibitem[{\citenamefont{{Ac{\'{\i}}n} and Masanes}(2016)}]{acin16}
\bibinfo{author}{\bibfnamefont{A.}~\bibnamefont{{Ac{\'{\i}}n}}}
  \bibnamefont{and} \bibinfo{author}{\bibfnamefont{L.}~\bibnamefont{Masanes}},
  \bibinfo{journal}{Nature} \textbf{\bibinfo{volume}{540}},
  \bibinfo{pages}{213} (\bibinfo{year}{2016}).

\bibitem[{\citenamefont{Bell}(1964)}]{Bell_original}
\bibinfo{author}{\bibfnamefont{J.~S.} \bibnamefont{Bell}},
  \bibinfo{journal}{Physics} \textbf{\bibinfo{volume}{1}}, \bibinfo{pages}{195}
  (\bibinfo{year}{1964}).

\bibitem[{\citenamefont{{Brunner} et~al.}(2014)\citenamefont{{Brunner},
  {Cavalcanti}, {Pironio}, {Scarani}, and {Wehner}}}]{Brunner:2014ix}
\bibinfo{author}{\bibfnamefont{N.}~\bibnamefont{{Brunner}}},
  \bibinfo{author}{\bibfnamefont{D.}~\bibnamefont{{Cavalcanti}}},
  \bibinfo{author}{\bibfnamefont{S.}~\bibnamefont{{Pironio}}},
  \bibinfo{author}{\bibfnamefont{V.}~\bibnamefont{{Scarani}}},
  \bibnamefont{and} \bibinfo{author}{\bibfnamefont{S.}~\bibnamefont{{Wehner}}},
  \bibinfo{journal}{Reviews of Modern Physics} \textbf{\bibinfo{volume}{86}},
  \bibinfo{pages}{419} (\bibinfo{year}{2014}), \eprint{1303.2849}.

\bibitem[{\citenamefont{{Christensen} et~al.}(2013)\citenamefont{{Christensen},
  {McCusker}, {Altepeter}, {Calkins}, {Gerrits}, {Lita}, {Miller}, {Shalm},
  {Zhang}, {Nam} et~al.}}]{Kwiat13}
\bibinfo{author}{\bibfnamefont{B.~G.} \bibnamefont{{Christensen}}},
  \bibinfo{author}{\bibfnamefont{K.~T.} \bibnamefont{{McCusker}}},
  \bibinfo{author}{\bibfnamefont{J.~B.} \bibnamefont{{Altepeter}}},
  \bibinfo{author}{\bibfnamefont{B.}~\bibnamefont{{Calkins}}},
  \bibinfo{author}{\bibfnamefont{T.}~\bibnamefont{{Gerrits}}},
  \bibinfo{author}{\bibfnamefont{A.~E.} \bibnamefont{{Lita}}},
  \bibinfo{author}{\bibfnamefont{A.}~\bibnamefont{{Miller}}},
  \bibinfo{author}{\bibfnamefont{L.~K.} \bibnamefont{{Shalm}}},
  \bibinfo{author}{\bibfnamefont{Y.}~\bibnamefont{{Zhang}}},
  \bibinfo{author}{\bibfnamefont{S.~W.} \bibnamefont{{Nam}}},
  \bibnamefont{et~al.}, \bibinfo{journal}{Physical Review Letters}
  \textbf{\bibinfo{volume}{111}}, \bibinfo{eid}{130406} (\bibinfo{year}{2013}),
  \eprint{1306.5772}.

\bibitem[{\citenamefont{{Bierhorst} et~al.}(2017)\citenamefont{{Bierhorst},
  {Knill}, {Glancy}, {Mink}, {Jordan}, {Rommal}, {Liu}, {Christensen}, {Nam},
  and {Shalm}}}]{nist17}
\bibinfo{author}{\bibfnamefont{P.}~\bibnamefont{{Bierhorst}}},
  \bibinfo{author}{\bibfnamefont{E.}~\bibnamefont{{Knill}}},
  \bibinfo{author}{\bibfnamefont{S.}~\bibnamefont{{Glancy}}},
  \bibinfo{author}{\bibfnamefont{A.}~\bibnamefont{{Mink}}},
  \bibinfo{author}{\bibfnamefont{S.}~\bibnamefont{{Jordan}}},
  \bibinfo{author}{\bibfnamefont{A.}~\bibnamefont{{Rommal}}},
  \bibinfo{author}{\bibfnamefont{Y.-K.} \bibnamefont{{Liu}}},
  \bibinfo{author}{\bibfnamefont{B.}~\bibnamefont{{Christensen}}},
  \bibinfo{author}{\bibfnamefont{S.~W.} \bibnamefont{{Nam}}}, \bibnamefont{and}
  \bibinfo{author}{\bibfnamefont{L.~K.} \bibnamefont{{Shalm}}},
  \bibinfo{journal}{ArXiv e-prints}  (\bibinfo{year}{2017}),
  \eprint{1702.05178}.

\bibitem[{\citenamefont{{Ac{\'{\i}}n} et~al.}(2016)\citenamefont{{Ac{\'{\i}}n},
  {Pironio}, {V{\'e}rtesi}, and {Wittek}}}]{Acin:2016cl}
\bibinfo{author}{\bibfnamefont{A.}~\bibnamefont{{Ac{\'{\i}}n}}},
  \bibinfo{author}{\bibfnamefont{S.}~\bibnamefont{{Pironio}}},
  \bibinfo{author}{\bibfnamefont{T.}~\bibnamefont{{V{\'e}rtesi}}},
  \bibnamefont{and} \bibinfo{author}{\bibfnamefont{P.}~\bibnamefont{{Wittek}}},
  \bibinfo{journal}{\pra} \textbf{\bibinfo{volume}{93}}, \bibinfo{eid}{040102}
  (\bibinfo{year}{2016}), \eprint{1505.03837}.

\bibitem[{\citenamefont{{Colbeck} and {Kent}}(2011)}]{CK11}
\bibinfo{author}{\bibfnamefont{R.}~\bibnamefont{{Colbeck}}} \bibnamefont{and}
  \bibinfo{author}{\bibfnamefont{A.}~\bibnamefont{{Kent}}},
  \bibinfo{journal}{Journal of Physics A Mathematical General}
  \textbf{\bibinfo{volume}{44}}, \bibinfo{eid}{095305} (\bibinfo{year}{2011}),
  \eprint{1011.4474}.

\bibitem[{\citenamefont{{Nieto-Silleras}
  et~al.}(2014)\citenamefont{{Nieto-Silleras}, {Pironio}, and
  {Silman}}}]{NPS14}
\bibinfo{author}{\bibfnamefont{O.}~\bibnamefont{{Nieto-Silleras}}},
  \bibinfo{author}{\bibfnamefont{S.}~\bibnamefont{{Pironio}}},
  \bibnamefont{and} \bibinfo{author}{\bibfnamefont{J.}~\bibnamefont{{Silman}}},
  \bibinfo{journal}{New Journal of Physics} \textbf{\bibinfo{volume}{16}},
  \bibinfo{eid}{013035} (\bibinfo{year}{2014}), \eprint{1309.3930}.

\bibitem[{\citenamefont{{Bancal} et~al.}(2014)\citenamefont{{Bancal},
  {Sheridan}, and {Scarani}}}]{BSS14}
\bibinfo{author}{\bibfnamefont{J.-D.} \bibnamefont{{Bancal}}},
  \bibinfo{author}{\bibfnamefont{L.}~\bibnamefont{{Sheridan}}},
  \bibnamefont{and}
  \bibinfo{author}{\bibfnamefont{V.}~\bibnamefont{{Scarani}}},
  \bibinfo{journal}{New Journal of Physics} \textbf{\bibinfo{volume}{16}},
  \bibinfo{eid}{033011} (\bibinfo{year}{2014}), \eprint{1309.3894}.

\bibitem[{\citenamefont{Pearle}(1970)}]{chained}
\bibinfo{author}{\bibfnamefont{P.~M.} \bibnamefont{Pearle}},
  \bibinfo{journal}{Phys. Rev. D} \textbf{\bibinfo{volume}{2}},
  \bibinfo{pages}{1418} (\bibinfo{year}{1970}),
  \urlprefix\url{https://link.aps.org/doi/10.1103/PhysRevD.2.1418}.

\bibitem[{\citenamefont{Kim et~al.}(2006)\citenamefont{Kim, Florentino, and
  Wong}}]{Kim2006a}
\bibinfo{author}{\bibfnamefont{T.}~\bibnamefont{Kim}},
  \bibinfo{author}{\bibfnamefont{M.}~\bibnamefont{Florentino}},
  \bibnamefont{and} \bibinfo{author}{\bibfnamefont{F.~N.~C.}
  \bibnamefont{Wong}}, \bibinfo{journal}{Phys. Rev. A}
  \textbf{\bibinfo{volume}{73}} (\bibinfo{year}{2006}).

\bibitem[{\citenamefont{Wong et~al.}(2007)\citenamefont{Wong, Shapiro, and
  Kim}}]{Kim2006b}
\bibinfo{author}{\bibfnamefont{F.~N.~C.} \bibnamefont{Wong}},
  \bibinfo{author}{\bibfnamefont{J.~H.} \bibnamefont{Shapiro}},
  \bibnamefont{and} \bibinfo{author}{\bibfnamefont{T.}~\bibnamefont{Kim}},
  \bibinfo{journal}{Laser physics} \textbf{\bibinfo{volume}{16}}
  (\bibinfo{year}{2007}).

\bibitem[{\citenamefont{Fedrizzi et~al.}(2007)\citenamefont{Fedrizzi, Herbst,
  Poppe, Jennewein, and Zeilinger}}]{Zeilinger2007}
\bibinfo{author}{\bibfnamefont{A.}~\bibnamefont{Fedrizzi}},
  \bibinfo{author}{\bibfnamefont{T.}~\bibnamefont{Herbst}},
  \bibinfo{author}{\bibfnamefont{A.}~\bibnamefont{Poppe}},
  \bibinfo{author}{\bibfnamefont{T.}~\bibnamefont{Jennewein}},
  \bibnamefont{and}
  \bibinfo{author}{\bibfnamefont{A.}~\bibnamefont{Zeilinger}},
  \bibinfo{journal}{Optics Express} \textbf{\bibinfo{volume}{15}}
  (\bibinfo{year}{2007}).

\bibitem[{\citenamefont{Ljunggren and Tengner}(2005)}]{Ljunggren_PRA_fibers}
\bibinfo{author}{\bibfnamefont{D.}~\bibnamefont{Ljunggren}} \bibnamefont{and}
  \bibinfo{author}{\bibfnamefont{M.}~\bibnamefont{Tengner}},
  \bibinfo{journal}{Phys. Rev. A} \textbf{\bibinfo{volume}{72}}
  (\bibinfo{year}{2005}).

\bibitem[{\citenamefont{G{\'o}mez et~al.}(2016)\citenamefont{G{\'o}mez,
  G{\'o}mez, Gonz{\'a}lez, Ca{\~n}as, Barra, Delgado, Xavier, Cabello,
  Kleinmann, V{\'e}rtesi et~al.}}]{gomez2016device}
\bibinfo{author}{\bibfnamefont{E.~S.} \bibnamefont{G{\'o}mez}},
  \bibinfo{author}{\bibfnamefont{S.}~\bibnamefont{G{\'o}mez}},
  \bibinfo{author}{\bibfnamefont{P.}~\bibnamefont{Gonz{\'a}lez}},
  \bibinfo{author}{\bibfnamefont{G.}~\bibnamefont{Ca{\~n}as}},
  \bibinfo{author}{\bibfnamefont{J.~F.} \bibnamefont{Barra}},
  \bibinfo{author}{\bibfnamefont{A.}~\bibnamefont{Delgado}},
  \bibinfo{author}{\bibfnamefont{G.~B.} \bibnamefont{Xavier}},
  \bibinfo{author}{\bibfnamefont{A.}~\bibnamefont{Cabello}},
  \bibinfo{author}{\bibfnamefont{M.}~\bibnamefont{Kleinmann}},
  \bibinfo{author}{\bibfnamefont{T.}~\bibnamefont{V{\'e}rtesi}},
  \bibnamefont{et~al.}, \bibinfo{journal}{Physical Review Letters}
  \textbf{\bibinfo{volume}{117}}, \bibinfo{pages}{260401}
  (\bibinfo{year}{2016}).

\bibitem[{\citenamefont{{Lin} et~al.}(2017)\citenamefont{{Lin}, {Rosset},
  {Zhang}, {Bancal}, and {Liang}}}]{regularization}
\bibinfo{author}{\bibfnamefont{P.-S.} \bibnamefont{{Lin}}},
  \bibinfo{author}{\bibfnamefont{D.}~\bibnamefont{{Rosset}}},
  \bibinfo{author}{\bibfnamefont{Y.}~\bibnamefont{{Zhang}}},
  \bibinfo{author}{\bibfnamefont{J.-D.} \bibnamefont{{Bancal}}},
  \bibnamefont{and} \bibinfo{author}{\bibfnamefont{Y.-C.}
  \bibnamefont{{Liang}}}, \bibinfo{journal}{ArXiv e-prints}
  (\bibinfo{year}{2017}), \eprint{1705.09245}.

\bibitem[{\citenamefont{Collins and Gisin}(2004)}]{CollinsGisin}
\bibinfo{author}{\bibfnamefont{D.}~\bibnamefont{Collins}} \bibnamefont{and}
  \bibinfo{author}{\bibfnamefont{N.}~\bibnamefont{Gisin}},
  \bibinfo{journal}{Journal of Physics A: Mathematical and General}
  \textbf{\bibinfo{volume}{37}}, \bibinfo{pages}{1775} (\bibinfo{year}{2004}),
  \urlprefix\url{http://stacks.iop.org/0305-4470/37/i=5/a=021}.

\bibitem[{\citenamefont{Zhang et~al.}(2011)\citenamefont{Zhang, Glancy, and
  Knill}}]{zhang11y}
\bibinfo{author}{\bibfnamefont{Y.}~\bibnamefont{Zhang}},
  \bibinfo{author}{\bibfnamefont{S.}~\bibnamefont{Glancy}}, \bibnamefont{and}
  \bibinfo{author}{\bibfnamefont{E.}~\bibnamefont{Knill}},
  \bibinfo{journal}{Phys. Rev. A} \textbf{\bibinfo{volume}{84}},
  \bibinfo{pages}{062118} (\bibinfo{year}{2011}).

\bibitem[{\citenamefont{{Wiseman} et~al.}(2007)\citenamefont{{Wiseman},
  {Jones}, and {Doherty}}}]{WJD07}
\bibinfo{author}{\bibfnamefont{H.~M.} \bibnamefont{{Wiseman}}},
  \bibinfo{author}{\bibfnamefont{S.~J.} \bibnamefont{{Jones}}},
  \bibnamefont{and} \bibinfo{author}{\bibfnamefont{A.~C.}
  \bibnamefont{{Doherty}}}, \bibinfo{journal}{Physical Review Letters}
  \textbf{\bibinfo{volume}{98}}, \bibinfo{eid}{140402} (\bibinfo{year}{2007}),
  \eprint{quant-ph/0612147}.

\bibitem[{\citenamefont{{Passaro} et~al.}(2015)\citenamefont{{Passaro},
  {Cavalcanti}, {Skrzypczyk}, and {Ac{\'{\i}}n}}}]{PCSA15}
\bibinfo{author}{\bibfnamefont{E.}~\bibnamefont{{Passaro}}},
  \bibinfo{author}{\bibfnamefont{D.}~\bibnamefont{{Cavalcanti}}},
  \bibinfo{author}{\bibfnamefont{P.}~\bibnamefont{{Skrzypczyk}}},
  \bibnamefont{and}
  \bibinfo{author}{\bibfnamefont{A.}~\bibnamefont{{Ac{\'{\i}}n}}},
  \bibinfo{journal}{New Journal of Physics} \textbf{\bibinfo{volume}{17}},
  \bibinfo{eid}{113010} (\bibinfo{year}{2015}), \eprint{1504.08302}.

\bibitem[{\citenamefont{{Navascu{\'e}s}
  et~al.}(2007)\citenamefont{{Navascu{\'e}s}, {Pironio}, and
  {Ac{\'{\i}}n}}}]{NPA}
\bibinfo{author}{\bibfnamefont{M.}~\bibnamefont{{Navascu{\'e}s}}},
  \bibinfo{author}{\bibfnamefont{S.}~\bibnamefont{{Pironio}}},
  \bibnamefont{and}
  \bibinfo{author}{\bibfnamefont{A.}~\bibnamefont{{Ac{\'{\i}}n}}},
  \bibinfo{journal}{Physical Review Letters} \textbf{\bibinfo{volume}{98}},
  \bibinfo{eid}{010401} (\bibinfo{year}{2007}), \eprint{quant-ph/0607119}.

\end{thebibliography}

\pagebreak

\section{\large Supplemental Material}

\setcounter{equation}{0}
\setcounter{figure}{0}
\setcounter{table}{0}
\setcounter{page}{1}
\makeatletter
\renewcommand{\theequation}{S\arabic{equation}}
\renewcommand{\thefigure}{S\arabic{figure}}
\renewcommand{\bibnumfmt}[1]{[S#1]}
\renewcommand{\citenumfont}[1]{S#1}

\maketitle
\section{Optical Implementation of a 3-element POVM}

\begin{figure}[h]
\centering
\includegraphics[width = .60\linewidth]{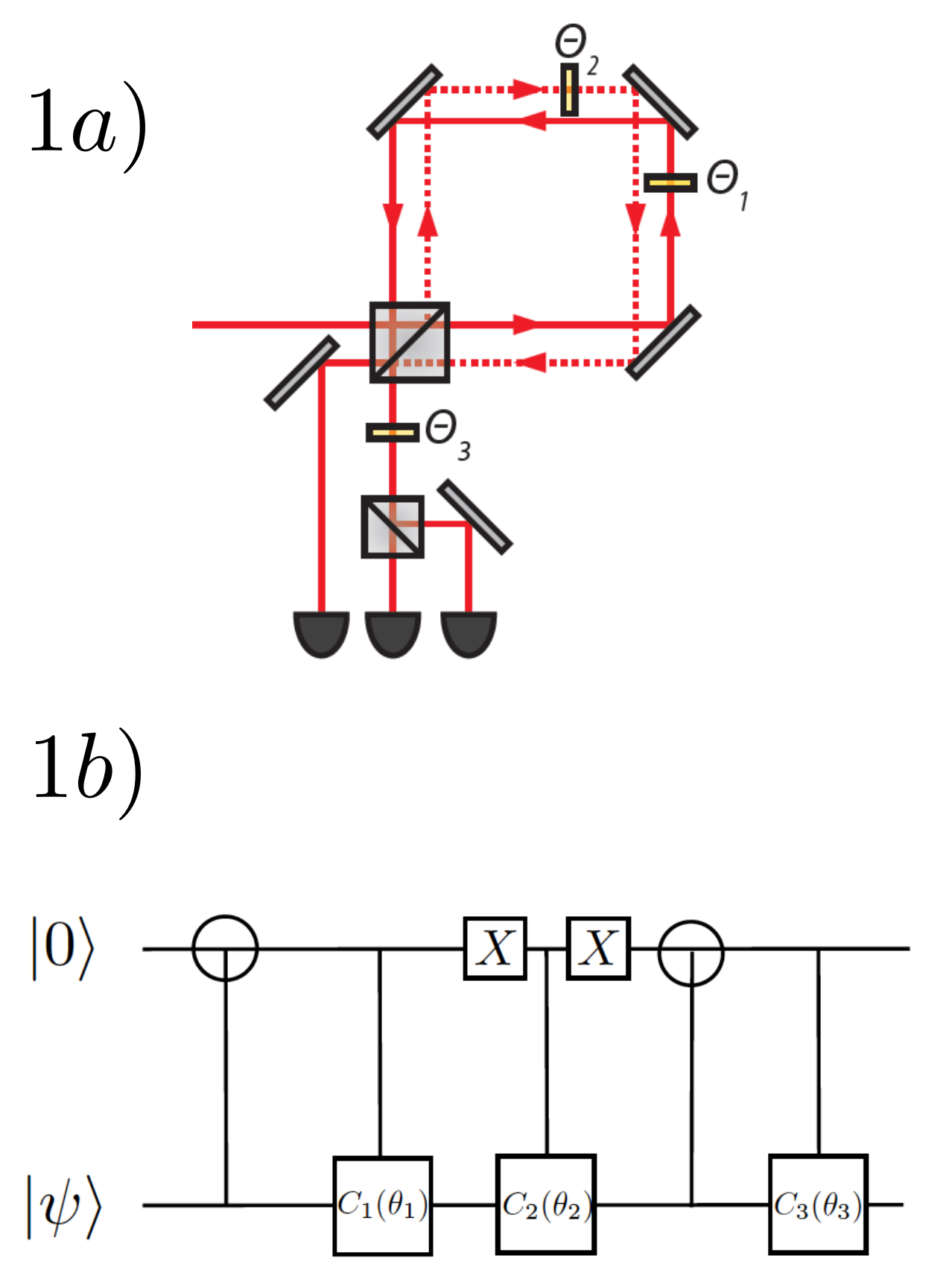}
\caption{In fig. $1a)$ the Sagnac interferometer used to implement the 3-outcome POVM. Fig. $1b)$ shows the corrersponding quantum circuit. }
\label{figSup}
\end{figure}

In order to implement a generalized measurement of more than two outcomes in the 2 dimensional polarization degree of freedom, we adopt the strategy of coupling to additional spatial modes through the Sagnac intereferometer in fig \ref{figSup}a).

We consider the polarization basis for the single photon $\{\ket{H},\ket{V}\}$ and two spatial modes $\{\ket{0},\ket{1}\}$ created by the Polarizing Beam Splitter (PBS) which defines the Sagnac interferometer.  The action of the interferometer can be better understood by the quantum circuit in fig \ref{figSup} b): A single photon with polarization $\ket{\psi }$, enters the through port $\ket{0}$ of the PBS and populates the spatial modes according to
\begin{eqnarray}
\ket{H}\ket{0} \rightarrow \ket{H}\ket{0} \nonumber \\
\ket{V}\ket{0} \rightarrow \ket{V}\ket{1}
\end{eqnarray}
The input port $\ket{1}$ is fed with vacuum, but an analogous analysis of its working completes de specification of the $CNOT$ gate implemented by the PBS. Once inside the interferometer, half wave plates at angles $\theta_1$ and $\theta_2$ rotate polarization depending on the spatial mode the photon. That is, internal half wave plates implement controlled operations, $C_1(\theta_1)$ controlled by the state $\ket{0}$ and $C_2(\theta_2)$.
A new passage through the same PBS implements a second $CNOT$ gate, controlled by the polarization. Finally we insert a half wave plate on output mode $\ket{0}$ at $\theta_3$ implementing the controlled operation $C_3(\theta_3)$.

The total unitary transformation which couples polarization with spatial modes is given
\begin{eqnarray}
U= C_3(\theta_3) *CNOT *C_2(\theta_2)* C_1(\theta_1) *CNOT \label{tU}
\end{eqnarray}
The coupling matrix (\ref{tU}) followed by detection in spatial modes, defines a family of POVM's in polarization parametrized by $\theta_1$, $\theta_2$ and $\theta_3$ .  In order to obtain this extremal POVM we chose the settings $\theta_1=0, \theta_2=2\rm{sin}\sqrt{2/3}, \theta_3=\pi/ 2$. In this situation, our interferometer works as selective attenuator of the $\ket{V}$ component, the part that has been attenuated, goes to output mode $\ket{1}$. As an example, we  consider the effect of $U$ on the states $\{\ket{\psi_0},\ket{\psi_1},\ket{\psi_2}\}$ of Eq.3 in the main text:
\begin{eqnarray}
\begin{array}{l}
\ket{\psi_0}  \rightarrow  \sqrt{1/6}\ket{0}\ket{H}+\sqrt{1/6}\ket{0}\ket{V}+ \sqrt{2/3}\ket{1}\ket{V} \\
\ket{\psi_1}\rightarrow \sqrt{2/3}\ket{0}\ket{H}+\sqrt{1/6}\ket{0}\ket{V}+ \sqrt{1/6}\ket{1}\ket{V}  \\
\ket{\psi_2} \rightarrow \sqrt{1/6}\ket{0}\ket{H}+\sqrt{2/3}\ket{0}\ket{V}+ \sqrt{1/6}\ket{1}\ket{V}.
\end{array}
\end{eqnarray}

By inserting a PBS in the outcome mode $\ket{0}$ we obtain the three outcome ports with the measurement statistics defining the POVM $\{ \frac{2}{3}\ket{\psi_0}\bra{\psi_0}, \frac{2}{3}\ket{\psi_1}\bra{\psi_1},\frac{2}{3}\ket{\psi_2}\bra{\psi_2}\}$.

\section{Device-independent randomness certification}
In our experiment we obtain a collection of observed experimental frequencies $f(ab|xy)$. Retrieving the amount of randomness from this data is not straightforward because the probability distributions $P(ab|xy)$ obtained upon normalizing these frequencies are ill defined due to the finite statistics regime intrinsic of any implementation. For instance they do not satisfy the no-signaling conditions (satisfied in quantum mechanics) defined by $\sum_bP(ab|xy) = \sum_bP(ab|xy')$ (no-signaling from B to A) and $\sum_aP(ab|xy) = \sum_aP(ab|x'y)$  (no-signaling from A to B).

In order to circumvent this problem we use the following steps. From the experimental frequencies $f(ab|xy)$ we generated a set of no-signaling probability distributions $P_{NS}=\{P_{NS}(ab|xy)\}$ through the Collins-Gisin parametrization of the space of probabilities \cite{CollinsGisin}. By considering marginal probabilities $P(a|x,y=1)$ and $P(b|x=1,y)$, the Collins-Gisin representation enforces no-signaling constraints of $P$ by dropping all probabilities involving the last outcome of all measurements. For instance, for the POVM that has 3 outcomes, the probability $P_{NS}(a=3,b|xy)$ is implicitly set by imposing $P_{NS}(a=3,b|xy) = P(a=3|x,y=1) - P(a=2,b|xy) - P(a=1|xy)$.

With $P_{NS}$ we run the semidefinite (SDP) program proposed in Refs. \cite{NPS14,BSS14}, that provides an upper bound to the guessing probability \eqref{eq:randomnessSDP}:
\begin{align}
\label{eq:randomnessSDP}
&P_{guess}=\max_{\{P(abe|xy)\}} P(a=e|x^*)\\
&\textrm{such that}\nonumber\\
& P(ab|xy)=\sum_{e} P(a,b,e\vert x,y) \quad\forall ~a,b,x,y\\
&P(abe|xy)\geq0\quad\forall ~a,b,e,x,y \\
&\sum_{abe} P(abe|xy)=1 \quad\forall ~x,y\\
& \{P(a,b,e\vert x,y)\}_{a,b,e,x,y} \in \mathcal{Q}_2.
\end{align}
This expression gives the maximum probability that Eve's outcomes match Alice's, given that the distributions observed are marginals of a joint tripartite distribution with Eve (constraints (2) and (3)). The last constraint (4) means that the joint distributions lie into the set $\mathcal{Q}_2$, an outer approximation to the set of quantum probability distributions $\mathcal{Q}$ proposed in \cite{NPA}.

The solution of this SDP optimization provides a linear function $S(P)$ whose value is a lower bound on the amount of randomness of any set of distributions $P$. We finally rewrite $S$ in terms of expected values and use it to estimate the amount of randomness in our experiment. The errors of the recorded probabilities are calculated assuming fair samples from Poissonian distributions and Gaussian error propagation. We note that our statistical analysis considers the asymptotic limit of many experimental runs. A more detailed statistical methods considering finite statistic \cite{Pironio:2010kx,zhang11y,nist17} is beyond the scope of this work.

\section{Randomness certification in the steering scenario}

The steering scenario refers to the situation where the box A is still treated as black boxes with inputs $x$ and outputs $a$, while B is assumed to be able to make tomography  of the conditional states $\rho_{a|x}^B$. The information the user has in this situation can be summarized in the set of unnormalized quantum sates  $\{\sigma_{a|x}^B\}_{a,x}$, where $\sigma_{a|x}=p(a|x)\rho_{a|x}^B$. Given the knowledge of $\{\sigma_{a|x}\}_{a,x}$, Alice and Bob can estimate the amount of randomness in Alice's outcomes through the following semi-definite program \cite{PCSA15}:
\begin{align}\label{e:local guess}
&P_{guess}(x^*)=\max_{\{\sigma_{a|x}^e\}}\quad \tr\sum_e \sigma_{a=e|x^*}^e \\
&\text{such that} \nonumber \\
&\sum_e \sigma_{a|x}^e = \sigma_{a|x}  \quad\forall a,x, \\
&\sum_a \sigma_{a|x}^e = \sum_a \sigma_{a|x'}^e \quad\forall e,x,x',  \\
&\sigma_{a|x}^e \geq 0 \quad\forall a,x,e.
\end{align}

Once more, the solution of this program gives a linear function (a quantum steering inequality) of the experimental data that can be used to calculate a the guessing probability and appropriate errors.

\end{document}